\documentclass[12pt]{iopart}

\usepackage{braket}
\usepackage{delarray}
\usepackage{color}
\usepackage{amssymb}

\begin{document}

\title[The hydrogen atom perturbed by a 1d-SHO potential]{The hydrogen atom perturbed by a 1-dimensional Simple Harmonic Oscillator (1d-SHO) potential}

\author{C. Santamarina R\'ios$^a$, P. Rodr\'iguez Cacheda$^b$ and J.J. Saborido Silva$^a$}

\address{$^a$ Instituto Galego de Física de Altas Enerxías (IGFAE), Universidade de Santiago de Compostela, Santiago de Compostela, Spain\\
$^b$ Universidade de Santiago de Compostela, Santiago de Compostela, Spain}
\ead{cibran.santamarina@usc.es}

\begin{abstract}
The hydrogen atom perturbed by a constant 1-dimensional weak quadratic potential $\lambda z^2$ is solved at first-order perturbation theory using the eigenstates of the total angular momentum operator - the coupled basis. Physical applications of this result could be found, for example, in the study of a quadratic Zeeman effect weaker than fine-structure effects, or in a perturbation caused by instantaneous generalised van der Waals interactions.
\end{abstract}

%
\vspace{2pc}

\submitto{\jpb}

%
%
%

\section{Introduction}

The hydrogen atom and the 1-dimensional Simple Harmonic Oscillator (1d-SHO), i.e., a particle under the effect of a one-dimensional quadratic potential, are certainly among the most studied systems in quantum mechanics. In fact, the hydrogen atom was the first system for which Erwin Schr\"odinger solved his celebrated equation~\cite{Schrodinger26}. The Wave Mechanics approach proposed by Schr\"odinger permits to obtain the hydrogen energy levels accounting for fine-structure effects that can be derived from the  Dirac equation in the non-relativistic regime with the use of perturbation theory. In the coupled basis, which describes the state of the system with the principal ($n$), total angular momentum ($j$), third component of the total angular momentum ($m$), and orbital angular momentum ($\ell$) quantum numbers, the energy eigenvalues for a hydrogen-like atom with a nucleus of charge $Z$ are given in Rydberg units by (see for instance~\cite{Weissbluth78})
\begin{equation} \label{eq:Hstrfn}
E_{nj}  = -\frac{Z^2}{n^2} \left[ 1 + \frac{Z^2\alpha^2}{n} \left( \frac{1}{j + \frac{1}{2}}
-\frac{3}{4n} \right) \right] \;\; {\rm Ry},
\end{equation}
where the fine-structure correction is ruled by the fine-structure constant squared ($\alpha^2$). Since this result only depends on $n$ and $j$ there remains a degeneracy of $2(2j+1)$, where the first factor of $2$ accounts for the spin degree of freedom.

The degenerate hydrogen levels with the same values of $n$, $j$ and $m$ can be represented, see for instance~\cite{Capri02}, by two multiplets,
\begin{eqnarray}
\fl
\qquad \ket{\phi_1} &\;=\; \frac{R_{n,j-\frac{1}{2}}(r)}{\sqrt{2j}}
\begin{array}({c}) \sqrt{j+m}\; Y_{j-\frac{1}{2}}^{m-\frac{1}{2}} \\
\sqrt{j-m}\; Y_{j-\frac{1}{2}}^{m+\frac{1}{2}} \end{array}, \hspace{22mm} \ell=j-\frac{1}{2}, \hspace{5mm}
\ell \ne 0, \\
\qquad \fl \ket{\phi_2} &\;=\; \frac{R_{n,j+\frac{1}{2}}(r)}{\sqrt{2(j+1)}}
\begin{array}({c}) \sqrt{j-m+1}\; Y_{j+\frac{1}{2}}^{m-\frac{1}{2}} \\
-\sqrt{j+m+1}\; Y_{j+\frac{1}{2}}^{m+\frac{1}{2}} \end{array}, \hspace{10mm} \ell=j+\frac{1}{2}, \hspace{5mm}
\ell \ne 0,
\end{eqnarray}
where $Y_\ell^m \equiv Y_\ell^m(\theta,\varphi)$ are spherical harmonics (the polar and azimuthal angle dependence is omitted to simplify the notation) and $R_{n,\ell}(r)$ are the radial functions. Since the electron spin is $\frac{1}{2}$, $\ell$ can either be $\ell=j-\frac{1}{2}$, in the case of $\ket{\phi_1}$, or $\ell=j+\frac{1}{2}$ in the case of $\ket{\phi_2}$. An exception occurs when $\ell=0$, because in that case $j$ takes a single value, $j=\frac{1}{2}$, leading to one doublet of states
\begin{equation}
\ket{\phi} = R_{n,0}(r)\,Y^0_0(\theta,\varphi)\, \begin{array}({c})
m+\frac{1}{2} \\ \frac{1}{2}-m \end{array}, \hspace{10mm} j=\frac{1}{2}, \hspace{5mm} \ell = 0.
\label{eq:l0states}
\end{equation} 

The system can be now placed under the influence of an additional interaction that, in order to be treated perturbatively, must be of less intensity than that of the fine-structure effects. An example of this is the perturbation by a linear potential, $V = \lambda z$, which can correspond, for example, to a weak-field Stark effect. This is addressed in~\cite{Capri02}, resulting in energy shifts of
\begin{equation}
\label{eq:linear_result}
\Delta E^{\rm L}_{njm} = \pm \frac{\lambda a_0}{Z} \frac{3n}{4}\,\frac{m \sqrt{n^2-(j+\frac{1}{2})^2}}{j(j+1)},
\end{equation}
where $a_0$ is the atom Bohr radius and the superindex L refers to the linear character of the potential.

In this work, the focus is on the perturbation by a squared potential, $V^{\rm Q}=\lambda z^2$, where the Q superindex stands for quadratic. The study has the particular appeal of combining the hydrogen atom and the 1d-SHO into one system. The final result can be combined with (\ref{eq:linear_result}) to solve a perturbation by a displaced 1d-SHO (DQ), $V^{\rm DQ}=\lambda (z-z_0)^2$.

The study of hydrogen under the influence of a quadratic potential has been previously performed using perturbation theory in the framework of the quadratic Zeeman or diamagnetic effects~\cite{Schiff38}, but under the assumption that its influence is stronger than that of the fine-structure effect and, consequently, employing the uncoupled angular basis. Therefore, the calculation presented here is of limited use in the strong-field scenario. However, quadratic effects also arise in the study of van der Waals forces~\cite{Lennard-Jones32,Cohen77,Ganesan90,Aspect02,deMelo13}. In this situation the calculation presented here can be useful, since the fine-structure effect is stronger than that of the external perturbation.

\section{Setting up the calculation}

This is a degenerate perturbation theory calculation that involves two states, $\ket{\phi_1}$ and $\ket{\phi_2}$, with the same $j$ quantum number, but different values of $\ell$. Matrix elements involving different $m$ states cancel. This can be seen by rewriting the potential as
\begin{equation}
V^{\rm Q} = \lambda z^2 = \lambda r^2 \cos^2\theta = \lambda \frac{2\sqrt{\pi}}{3} r^2 \left(\frac{2}{\sqrt{5}}Y_2^0+Y_0^0\right).
\end{equation}
Since only $m=0$ spherical harmonics contribute to $V^{\rm Q}$, the Wigner-Eckart theorem ensures that $m$ is conserved in the matrix elements of this potential. Moreover, since $V^{\rm Q}=\lambda z^2$ is an even-parity function and $\ket{\phi_1}$ and $\ket{\phi_2}$ have different parities, the off-diagonal matrix elements are zero: $\Braket{\phi_1|V^{\rm Q}|\phi_2} = \Braket{\phi_2|V^{\rm Q}|\phi_1} =0$. Therefore, finding the $2 \times 2$ matrix eigenvalues of $V^{\rm Q}$ in the sub-space generated by $\ket{\phi_1}$ and $\ket{\phi_2}$ is reduced to compute $\Braket{\phi_1|V^{\rm Q}|\phi_1}$ and $\Braket{\phi_2|V^{\rm Q}|\phi_2}$, which will directly provide the energy shifts produced by $V^{\rm Q}$ in the first order of perturbation theory.

\section{Calculating the matrix elements of the the squared potential}
\label{sec:squarematrix}

The first energy shift is given by
\begin{eqnarray} \nonumber \fl
\Braket{\phi_1|V^{\rm Q}|\phi_1}= \int \int
\frac{R_{n,j-\frac{1}{2}}(r)}{\sqrt{2j}}
\left(
\sqrt{j+m}\;\left(Y_{j-\frac{1}{2}}^{m-\frac{1}{2}}\right)^\ast,
\sqrt{j-m}\;\left(Y_{j-\frac{1}{2}}^{m+\frac{1}{2}}\right)^\ast
\right) \\ \hspace{-15mm} \times\;
\lambda \frac{2\sqrt{\pi}}{3} r^2 \left(\frac{2}{\sqrt{5}}Y_2^0+Y_0^0\right)
\frac{R_{n,j-\frac{1}{2}}(r)}{\sqrt{2j}}
\left(
\begin{array}{c}
\sqrt{j+m}\;Y_{j-\frac{1}{2}}^{m-\frac{1}{2}} \\
\sqrt{j-m}\;Y_{j-\frac{1}{2}}^{m+\frac{1}{2}}
\end{array}
\right) r^2\sin\theta\, \rmd\,\varphi\,\rmd\theta\,\rmd\,r.
\end{eqnarray}
After some algebra, and using the Dirac braket notation for the angular integral,
\begin{eqnarray} \nonumber \fl
\Braket{\phi_1|V^{\rm Q}|\phi_1} = \lambda\frac{\sqrt{\pi}}{3j}\int_0^\infty R^2_{n,j-\frac{1}{2}}(r)\,r^4\, {\rm d}r \\
\nonumber \times
\left(\frac{2(j+m)}{\sqrt{5}}\Braket{Y_{j-\frac{1}{2}}^{m-\frac{1}{2}}|Y_2^0|Y_{j-\frac{1}{2}}^{m-\frac{1}{2}}}
+\frac{2(j-m)}{\sqrt{5}}\Braket{Y_{j-\frac{1}{2}}^{m+\frac{1}{2}}|Y_2^0|Y_{j-\frac{1}{2}}^{m+\frac{1}{2}}}\right.\\
\left. +\;(j+m)\Braket{Y_{j-\frac{1}{2}}^{m-\frac{1}{2}}|Y_0^0|Y_{j-\frac{1}{2}}^{m-\frac{1}{2}}}
+(j-m)\Braket{Y_{j-\frac{1}{2}}^{m+\frac{1}{2}}|Y_0^0|Y_{j-\frac{1}{2}}^{m+\frac{1}{2}}}\right),
\label{eq:phi1matrixel}
\end{eqnarray}
where the radial integral is a well-known calculation that can be found in almost any Quantum Mechanics textbook, for instance~\cite{Weissbluth78,Bethe57}, corresponding to the radial average of $r^2$,
\begin{equation}
\Braket{r^2}_{n,\ell}= \int_0^\infty R^2_{n,\ell}(r)\,r^4\, {\rm d}r = \frac{a_0^2}{Z^2}\frac{n^2}{2}\left[5n^2+1-3\ell(\ell+1)\right].
\label{eq:radial}
\end{equation}

Matrix elements containing three spherical harmonics can be calculated with the Gaunt integral~\cite{Gaunt29}. The result quoted here is expressed in terms of Wigner 3j-coefficients, following~\cite{Weissbluth78},
\begin{equation}
\fl
\braket{Y_{\ell^\prime}^{m^\prime}|Y_L^M|Y_\ell^m} =
(-1)^{m^\prime}\sqrt{\frac{(2\ell^\prime+1)(2L+1)(2\ell+1)}{4\pi}}
\left( \begin{array}{ccc} \ell^\prime & L & \ell \\ -m^\prime & M & m \end{array} \right)
\left( \begin{array}{ccc} \ell^\prime & L & \ell \\ 0 & 0 & 0 \end{array} \right).
\end{equation}
Using this expression, the contributions in~(\ref{eq:phi1matrixel}) containing the $Y_2^0$ spherical harmonic can be written as
\begin{eqnarray} \nonumber \fl
\sqrt{\frac{4\pi}{5}}\frac{1}{3j}\left[
(j+m)\Braket{Y_{j-\frac{1}{2}}^{m-\frac{1}{2}}|Y_2^0|Y_{j-\frac{1}{2}}^{m-\frac{1}{2}}}+
(j-m)\Braket{Y_{j-\frac{1}{2}}^{m+\frac{1}{2}}|Y_2^0|Y_{j-\frac{1}{2}}^{m+\frac{1}{2}}}\right] = \\ \nonumber
\left(\frac{2}{3}\begin{array}{ccc} j-\frac{1}{2} & 2 & j-\frac{1}{2} \\ 0 & 0 & 0 \end{array}\right)
\Bigg[ (-1)^{m-\frac{1}{2}}(j+m)
\left(\begin{array}{ccc} j-\frac{1}{2} & 2 & j-\frac{1}{2} \\ -(m-\frac{1}{2}) & 0 & m-\frac{1}{2} \end{array}\right) \\
+\;(-1)^{m+\frac{1}{2}}(j-m) \left(\begin{array}{ccc} j-\frac{1}{2} & 2 & j-\frac{1}{2} \\
-(m+\frac{1}{2}) & 0 & m+\frac{1}{2} \end{array}\right) \Bigg].
\label{eq:Y20_first}
\end{eqnarray}
The 3j-Wigner symbols in this equation can be obtained, see for instance,~\cite{Edmonds95}, with a tabulated result
\begin{equation}
\left( \begin{array}{ccc} \ell & 2 & \ell \\ -m & 0 & m \end{array}
\right)=(-1)^{\ell-m} \frac{2[3m^2-\ell(\ell+1)]}{\sqrt{(2\ell+3)(2\ell+2)(2\ell+1)(2\ell)(2\ell-1)}}.
\end{equation}
Using this expression it is found that
\begin{equation}
\left(\begin{array}{ccc} j-\frac{1}{2} & 2 & j-\frac{1}{2} \\ 0 & 0 & 0 \end{array}\right) =
\frac{(-1)^{j+\frac{1}{2}}}{2\sqrt{2}}\; \sqrt{\frac{(j+\frac{1}{2})(j-\frac{1}{2})}{(j+1)j(j-1)}},
\end{equation}
\begin{equation}
\fl \qquad
\left(\begin{array}{ccc} j-\frac{1}{2} & 2 & j-\frac{1}{2} \\ -(m+\frac{1}{2}) & 0 & (m+\frac{1}{2}) \end{array}\right) =
\frac{(-1)^{j-m-1}}{2\sqrt{2}}\; \frac{3(m+\frac{1}{2})^2-(j-\frac{1}{2})(j+\frac{1}{2})}
{\sqrt{(j+1)(j+ \frac{1}{2})j(j-\frac{1}{2})(j-1)}}.
\end{equation}
With this, and after some simplifications, (\ref{eq:Y20_first}) reduces to
\begin{eqnarray} \nonumber \fl
\sqrt{\frac{4\pi}{5}}\frac{1}{3j}\left[
(j+m)\Braket{Y_{j-\frac{1}{2}}^{m-\frac{1}{2}}|Y_2^0|Y_{j-\frac{1}{2}}^{m-\frac{1}{2}}}+
(j-m)\Braket{Y_{j-\frac{1}{2}}^{m+\frac{1}{2}}|Y_2^0|Y_{j-\frac{1}{2}}^{m+\frac{1}{2}}}\right]
\\= -\frac{1}{6} \frac{3m^2-j(j+1)}{(j+1)j},
\label{assig2-2023-1}
\end{eqnarray}
where it was used that $(-1)^{2j}=-1$, because $2j$ is always an odd number, $2j = 2\ell \pm 1$.

The terms containing $Y^0_0 = 1/\sqrt{4\pi}$ are easy to compute using the orthonormality condition of the spherical harmonics
\begin{equation} \label{assig2-2023-2}
\fl \qquad
\frac{\sqrt{\pi}}{3j}\left[(j+m)\Braket{Y_{j-\frac{1}{2}}^{m-\frac{1}{2}}|Y_0^0|Y_{j-\frac{1}{2}}^{m-\frac{1}{2}}}
+(j-m)\Braket{Y_{j-\frac{1}{2}}^{m+\frac{1}{2}}|Y_0^0|Y_{j-\frac{1}{2}}^{m+\frac{1}{2}}}\right] = \frac{1}{3}.
\end{equation}
With this result,
\begin{equation}
\Braket{\phi_1|V^{\rm Q}|\phi_1} =
\lambda \Braket{r^2}_{n,j-\frac{1}{2}}\frac{1}{2}\left(1 - \frac{m^2}{(j+1)j} \right).
\label{eq:phi1matrixelr}
\end{equation}
The calculation of $\Braket{\phi_2|V^{\rm Q}|\phi_2}$ proceeds similarly. After defining $j'=j+1$, an analogue to (\ref{eq:phi1matrixel}) is found,
\begin{eqnarray} \nonumber \fl
\Braket{\phi_2|V^{\rm Q}|\phi_2} =  \lambda\frac{\sqrt{\pi}}{3j'}\int_0^\infty R^2_{n,j'-\frac{1}{2}}(r)\,r^4\, {\rm d}r \\
\nonumber \times\; \biggl(
 \frac{2(j'-m)}{\sqrt{5}}\Braket{Y_{j'-\frac{1}{2}}^{m-\frac{1}{2}}|Y_2^0|Y_{j'-\frac{1}{2}}^{m-\frac{1}{2}}}
+\frac{2(j'+m)}{\sqrt{5}}\Braket{Y_{j'-\frac{1}{2}}^{m+\frac{1}{2}}|Y_2^0|Y_{j'-\frac{1}{2}}^{m+\frac{1}{2}}} \\
+\;(j'-m)\Braket{Y_{j'-\frac{1}{2}}^{m-\frac{1}{2}}|Y_0^0|Y_{j'-\frac{1}{2}}^{m-\frac{1}{2}}}
+(j'+m)\Braket{Y_{j'-\frac{1}{2}}^{m+\frac{1}{2}}|Y_0^0|Y_{j'-\frac{1}{2}}^{m+\frac{1}{2}}}
\biggl).
\end{eqnarray}
This expression contains the same spherical harmonic brakets and radial integral computed for (\ref{eq:phi1matrixel}) and discussed above. A very similar calculation leads to,
\begin{equation}
\fl \;\;
\Braket{\phi_2|V^{\rm Q}|\phi_2} = \lambda
\Braket{r^2}_{n,j'-\frac{1}{2}}\frac{1}{2}\left( 1 - \frac{m^2}{j'(j'-1)} \right)=\lambda
\Braket{r^2}_{n,j+\frac{1}{2}}\frac{1}{2}\left( 1 - \frac{m^2}{(j+1)j} \right),
\label{eq:phi2matrixelr}
\end{equation}
where the $j'=j+1$ change was undone in the last step.
Using the orbital angular momentum quantum number $\ell$ and~(\ref{eq:radial}), the two results of~(\ref{eq:phi1matrixelr}) and~(\ref{eq:phi2matrixelr}), can be synthesised in a general expression applicable to any $\ket{n\ell j m}$ state of the coupled basis, which also provides the energy shift of the quadratic (Q) 1d-SHO potential,
\begin{eqnarray} \nonumber
\Delta E^{\rm Q}_{n\ell j m} & = \Braket{n \ell j m|V^{\rm Q}|n\ell j m} = 
\lambda \Braket{r^2}_{n,\ell}\frac{1}{2}
\left(1 - \frac{m^2}{(j+1)j} \right)
\\ &= \lambda \left(\frac{a_0n}{2Z}\right)^2\left[5n^2+1-3\ell(\ell+1)\right]
\left(1 - \frac{m^2}{(j+1)j} \right),
\label{eq:eqresult}
\end{eqnarray}
which is the main result of this paper.  It is not difficult to check that this expression remains also valid for the $\ell=0$ states, given by~(\ref{eq:l0states}).

\section{Generalising the result to a displaced 1d-SHO}

The results obtained for the perturbation of the hydrogen atom with linear and quadratic potentials can be combined to compute the energy shifts produced by a perturbation of the form
\begin{equation}
V^{\rm DQ} =\lambda(z-z_0)^2=\lambda z^2-2\lambda z_0 z + \lambda z_0^2,
\label{eq:dqpotential}
\end{equation}
where the label DQ stands for {\em displaced quadratic}. As for the linear potential, this perturbation does not commute with parity and, therefore, will mix even and odd parity eigenstates. Thus, $\ell$ will not be a good quantum number for the energy shifts.

However, $m$ is a good quantum number and, similarly to the discussion of section~\ref{sec:squarematrix}, the calculation can be restricted to two-dimensional spaces generated by the $\ket{\phi_1}$ and $\ket{\phi_2}$ states. The general solution to the diagonalisation of a $2 \times 2$ hermitian operator,
\[ \mathcal{H} = \left( \begin{array}{cc} \mathcal{H}_{11} & \mathcal{H}_{12} \\
\mathcal{H}^\ast_{12} & \mathcal{H}_{22} \\ \end{array} \right), \]
is given by the eigenvalues,
\begin{equation} 
\lambda_\pm = \frac{1}{2} \left[\mathcal{H}_{11}+\mathcal{H}_{22}\pm
\sqrt{(\mathcal{H}_{11}-\mathcal{H}_{22})^2+4|\mathcal{H}_{12}|^2}\right].
\label{eq:eigenvalues}
\end{equation}
For the potential~(\ref{eq:dqpotential}) the constant contribution, $\lambda z_0^2$, produces a global shift that can be neglected in the calculation and added to the final result. Due to parity conservation, discussed also above, the quadratic term only contributes to the diagonal elements, whereas the linear terms only contribute to the off-diagonal elements. Therefore, $\mathcal{H}_{11}$ is given by~(\ref{eq:phi1matrixelr}), $\mathcal{H}_{22}$ is given by~(\ref{eq:phi2matrixelr}) and $|\mathcal{H}_{12}|^2=4z_0^2\left(\Delta E_{njm}^{\rm L}\right)^2$, where $\Delta E_{njm}^{\rm L}$ was quoted in~(\ref{eq:linear_result}). Plugging-in these values in~(\ref{eq:eigenvalues}) and accounting for the global shift, the energy corrections are found to be
\begin{eqnarray} \nonumber \fl
\Delta E^{\rm DQ}_{n j m} = \lambda z_0^2 + \lambda_\pm = \lambda z_0^2 + \lambda \left(\frac{a_0n}{2Z}\right)^2
\left(1 - \frac{m^2}{(j+1)j} \right) \Bigg\{5n^2-3j(j+1)+\frac{1}{4} \\ \pm
3 (j+\textstyle\frac{1}{2}\displaystyle)\left[1+\left(\frac{2Zz_0}{a_0}\frac{m}{(j(j+1)-m^2)}\right)^2\left(\frac{1}{(j+\frac{1}{2})^2}-
\frac{1}{n^2}\right)\right]^{\frac{1}{2}}\Bigg\}.
\end{eqnarray}

\section{Using the result with the instantaneous generalised van der Waals interaction}

Another case of interest to employ the result of~(\ref{eq:eqresult}) is the perturbation caused by an instantaneous generalised van der Waals interaction potential, which is given by~\cite{Ganesan90},
\begin{equation}
V^{\rm GW} = \gamma (x^2+y^2+\beta^2z^2)=\gamma (r^2+(\beta^2-1)z^2).
\end{equation}
The energy shift can be easily computed identifying $\lambda = \gamma(\beta^2-1)$, obtaining
\begin{equation}
\Delta E^{\rm GW}_{n\ell jm} = \gamma \braket{r^2}_{n,\ell} + \Delta E^{\rm Q}_{n\ell j m}= \frac{\gamma}{2} \braket{r^2}_{n,\ell}\left[1+\beta^2+(1-\beta^2)\frac{m^2}{j(j+1)}\right].
\label{eq:vanderWaals}
\end{equation}
A particular case is the Lennard-Jones potential~\cite{Lennard-Jones32} acting on a hydrogen atom ($Z=1$),\footnote{CGS units are employed.}
\begin{equation}
V^{\rm LJ} = -\frac{e^2}{16d^3} \left(x^2+y^2+2z^2\right),
\end{equation}
which represents the interaction of the atom with an ideal infinite conducting wall placed at a distance $d$. For this potential $\gamma=-\frac{e^2}{16d^3}$ and $\beta^2=2$, and the result of (\ref{eq:vanderWaals}) is,
\begin{equation}
\Delta E^{\rm LJ}_{n\ell jm} = -\left(\frac{a_0}{2d}\right)^3\frac{n^2}{2}\left[5n^2+1-3\ell(\ell+1)\right] \left[3-\frac{m^2}{j(j+1)}\right] \;\;{\rm Ry}.
\label{eq:ljshift}
\end{equation}
An interesting result is that of the ground state $\Delta E^{\rm LJ}=-(a_0/d)^3\; {\rm Ry}$. Since the fine-structure effects have spherical symmetry, the result is also valid in the case of $V^{\rm LJ}$ being stronger than the former.

\section{Conclusions}

The energy shift of the hydrogen atom levels has been calculated for a one-dimensional quadratic perturbation (1d-SHO) in the coupled basis. This complements the calculation of linear perturbations in the same basis presented in~\cite{Capri02}. The expressions are generalised for the case of a displaced 1d-SHO and for the instantaneous generalised van der Waals interaction. The applicability of the obtained results is limited to cases without stronger perturbations that do not commute with the total angular momentum, which would imply the use of the uncoupled basis, for example the linear Zeeman effect. However, for atomic hydrogen, this is the case. Using the law of ideal gases it can be found that the volume per atom at pressure $P$ and temperature $T$ is
\[
d^3 \sim 40 \frac{P_0}{P}\frac{T}{T_0} \; {\rm nm}^3,
\]
where $P_0$ is the normal pressure of 101325 Pa and $T_0$ the normal temperature of 293.15 K. At normal pressure and temperature hydrogen appears in the ${\rm H}_2$ molecular state, whereas atomic ${\rm H}_1$ requires $1 \ll \frac{P_0}{P}\frac{T}{T_0}$. Thus, for atomic hydrogen $40 \; {\rm nm}^3 \ll d^3$ and in (\ref{eq:ljshift})
\begin{equation}
\label{eq:LJcondition}
\left(\frac{a_0}{2d}\right)^3 \ll 4.6\times 10^{-7},
\end{equation}
whereas the fine-structure correction is of the order of $\alpha^2 \sim 5.3 \times 10^{-5}$, fulfilling the hypothesis that the fine structure corrections are stronger than the perturbation, at least for the lowest $n$ states.\footnote{For excited states it has to be taken into account that, while fine-structure corrections go as $n^{-3}$ the result of (\ref{eq:ljshift}) goes as $n^4$, and the dominance of the fine-structure breaks down.} It is also true that for distances of the order of $d \sim c\tau\sim 0.3 \; {\rm m}$, where $\tau\sim 10^{-9} \; {\rm s}$ is the order of magnitude of the lifetime of the bound states, retardation effects arise and a more complex approach has to be considered~\cite{Aspect02}. 

The Lamb shift in $\ell=0$ hydrogen states is of the same order of magnitude than the Lennard-Jones potential fulfilling (\ref{eq:LJcondition})
\[ \Delta E^{\rm L} \sim 10^{-6}\frac{1}{n^3}\; {\rm Ry}, \]
whereas for $\ell \neq 0$ states
\[
\Delta E^{\rm L} \lesssim 10^{-9}\frac{1}{n^3}\; {\rm Ry}.
\]
However, the choice of the coupled basis is determined by the fine-structure interaction, which is more than two orders of magnitude stronger than the $\ell =0$ states Lamb shift and the correction originated in the quadratic potential. Therefore, the Lamb shift effect is simply to be added to the presented result.

A different scenario arises if the hyperfine interaction is comparable to the effects produced by the quadratic potential, since the coupling to the nuclear spin can have significant effects and $m$ in (\ref{eq:eqresult}) would no longer be a good quantum number. In the case of the Lennard-Jones potential this would be avoided if
\[
\alpha^2 \frac{m_e}{m_p} \sim 2.9\times 10^{-8}  \ll \left(\frac{a_0}{2d}\right)^3,
\]
where $m_e$ and $m_p$ are the electron and proton mass, respectively.

\ack

The authors would like to thank Prof. N\'estor Armesto and Prof. Elena Gonz\'alez-Ferreiro for their careful reading of the manuscript. CSR and JJSS would also like to acknowledge their students of the {\em F\'isica Cu\'antica III} course at Universidade de Santiago de Compostela for motivating this work. This work has been partially funded by {\em Xunta de Galicia} through the ED431C 2022/30 research grant.

\section*{References}

\end{document}